\documentclass[reprint,
 amsmath,amssymb,
 aps, superscriptaddress, floatfix
]{revtex4-1}

\usepackage{graphicx}
\usepackage{hyperref}
\usepackage{dcolumn}
\usepackage{bm}
\usepackage{amsmath,amssymb,mathrsfs,physics,tikz}
\newcommand{\RN}[1]{%
\textup{\uppercase\expandafter{\romannumeral#1}}%
}
\usepackage{float}

\begin{document}

\setlength{\abovedisplayskip}{3pt}
\setlength{\belowdisplayskip}{6pt}
\setlength{\abovecaptionskip}{3pt}
\setlength{\belowcaptionskip}{6pt}


\title{Spatiotemporal Determinants of Vector-Borne Disease Outbreaks}

\author{Bibandhan Poudyal,}
\affiliation{Department of Physics \& Astronomy, University of Rochester, Rochester, NY, 14627, USA}
\author{Gourab Ghoshal}
\email{gghoshal@pas.rochester.edu}
\affiliation{Department of Physics \& Astronomy, University of Rochester, Rochester, NY, 14627, USA}
\affiliation{Department of Computer Science, University of Rochester, Rochester, NY, 14627, USA}


\begin{abstract}
Vector-borne diseases arise from the coupled dynamics of human mobility and mosquito ecology, producing outbreaks shaped by both spatial distributions and temporal patterns of movement. Here we develop a coarse-grained hub–leaf reduction that isolates the universal principles governing epidemic vulnerability in interconnected populations. By deriving and analyzing the epidemic vulnerability equation, we show how human and vector population ratios, together with mobility parameters that regulate time spent in hub and leaf locations, jointly determine the conditions for outbreak persistence. The analysis reveals that balanced flows of individuals between patches consistently minimize vulnerability, whereas disproportionate concentrations of vectors can shift the dominant risk to specific locations. Across parameter regimes, compensatory mobility emerges as a stabilizing mechanism, while skewed host–vector ratios elevate epidemic risk. These results establish general principles for the spatiotemporal determinants of vector-borne disease spread and provide a theoretical foundation for extending minimal models to more complex epidemiological settings.
\end{abstract}

\maketitle

\section{Introduction}
\vspace{0.3cm}

Vector-borne diseases (VBDs), primarily transmitted by insects such as mosquitoes, remain a major global health challenge \cite{us2024national}. The World Health Organization (WHO) reports that they are responsible for a substantial fraction of global illness, placing more than half of the world's population at risk \cite{world2004world}. The dynamics of VBD transmission emerge from the interplay of vector ecology, host behavior, and environmental factors \cite{reiter1996global,eder2018scoping,ma2022climate,brass2024role,hales2002potential,ryan2019global,athni2021influence,abbasi2025climate}. In recent decades, the rapid growth of human mobility across urban, regional, and long-distance scales has further amplified the spread of these diseases, contributing to numerous outbreaks worldwide \cite{barbosa2018human,lee2017morphology,gonzalez2008understanding}. Urban scaling of social ties connects population density to contact opportunities and contagion potential \cite{Pan_2013,connolly2021extended,schlapfer2014scaling}, while intercity mobility networks correlate with socioeconomic performance and exposure pathways \cite{Mimar_2022, duenas2021changes,ruan2015integrated}. These effects are further shaped by inequality and access to infrastructure, which modulate movement patterns across cities \cite{Barbosa_2021,spray2022inequitable}.

Because at-risk populations are interconnected through flows of people, mobility plays a pivotal role in the spatial spread of VBDs, much as it does in human-to-human transmission~\cite{panos_2022}. This role has been established in airborne epidemics through network-based studies \cite{colizza2006role,adams2009man,arenas2020modeling,belik2011natural,hazarie2021interplay, panos_2020} and extended to vector-borne contexts, where human travel links otherwise isolated patches \cite{cosner2009effects,soriano2020vector,wesolowski2012quantifying,prosper2012assessing}. Network and metapopulation models provide a natural framework for capturing this interplay, representing geographic locations as nodes and human flows as links. Colocation networks and information-theoretic analyses further show that mobility flows encode substantial predictive structure \cite{Chen_2022, Poudyal_2024}, while planar urban networks exhibit invariant flow backbones that motivate coarse-grained reductions to capture dominant pathways \cite{Kirkley_2018}. More broadly, reaction--diffusion dynamics on heterogeneous networks generate robust spatial patterning \cite{Mimar_2021}, highlighting how structural heterogeneity alone can shape emergent spreading dynamics. Incorporating mobility into such models is essential for predicting contagion patterns, identifying the drivers of epidemic vulnerability, and ultimately informing strategies to reduce the risk of VBD outbreaks.

Metapopulation models, originally developed in ecology \cite{tilman1997spatial,hanski1997metapopulation}, offer a powerful framework for integrating mobility and contagion dynamics by representing geographical locations as nodes (e.g., neighborhoods or cities) where large groups of individuals reside and interact, with links between these nodes quantifying individual travel \cite{grenfell1997meta,eubank2004modelling,colizza2007reaction,balcan2010modeling,xiao2014transmission,moulay2013metapopulation,soriano2018spreading}. Drawing inspiration from multi-patch approaches, this paper explores the parameter space of VBD spread within a simplified one-hub--leaf mobility network. While broader metapopulation frameworks can analyze epidemic vulnerability across many interconnected regions, focusing on this minimal spatial structure provides a principled starting point for isolating the mechanisms that govern transmission between a densely populated ``hub'' and a more sparsely populated ``leaf'' area. Such reductions resonate with broader approaches in network science, where simple generative or dynamical rules have been used to uncover universal properties of growth and resilience before extending to more elaborate settings \cite{Ghoshal_2007,Karrer_2008}.

Our previous study \cite{poudyal2025contrasting} demonstrated the value of this reductionist framework in separating spatial and temporal components of disease spread \cite{aguilar2022impact,belik2011natural,balcan2009multiscale,poletto2013human,blonder2012temporal,holme2016temporal,hazarie2021interplay,ruan2007spatial,lloyd2004spatiotemporal}. By collapsing complex mobility networks into a tractable hub--leaf representation, we were able to uncover the fundamental drivers of contagion dynamics and make the trade-offs of mobility-targeted interventions transparent. In particular, the framework clarified results from the city of Cali, Colombia, where contrasting outcomes for airborne and vector-borne diseases under identical mobility restrictions were difficult to interpret in a full metapopulation setting. In the present work, we extend this framework beyond intervention analysis to identify the universal spatiotemporal principles that determine vulnerability in vector-borne diseases.

Motivated by these insights, we also recognize a key limitation of our previous work \cite{poudyal2025contrasting}. Although longitudinal analyses revealed diverse vector population distributions over a decade, our earlier study confined the spatiotemporal investigation to a uniformly low vector-to-human ratio, approximately one mosquito per hundred people, because the primary focus was on mobility intervention strategies. This restriction left unexplored the broader parameter landscapes that shape epidemic vulnerability.

In this paper, we address this gap by systematically exploring the spatiotemporal parameter space of the one-hub--leaf VBD model. To guide this analysis, we categorize parameters according to their influence on spatial versus temporal aspects of disease spread. Spatial parameters include the human and mosquito population densities within the hub and the leaf, which directly affect local transmission rates and the availability of hosts and vectors. Temporal parameters capture the allocation of time individuals spend in each location, represented by the mobility value between hub and leaf, which shapes their exposure risk and transmission potential.

By systematically varying these key parameters---human and mosquito population densities (spatial) and mobility-related time allocation (temporal)---we map how epidemic vulnerability responds across parameter space, quantifying how changes in spatial and temporal factors alter both the likelihood and the severity of outbreaks. The insights gained from this parameter space exploration provide a foundation for understanding how spatial population distributions and temporal presence patterns govern epidemic vulnerability, and lay the groundwork for extending minimal reductions toward more complex models and targeted intervention strategies. We show that vulnerability to VBD outbreaks is governed by a small set of universal principles: the relative abundance of vectors, the distribution of human hosts, and the balance of mobility between densely and sparsely populated regions. In particular, we find that a balanced allocation of time between hub and leaf consistently minimizes vulnerability, while disproportionate vector concentrations strongly elevate risk. 

The paper is organized as follows. We begin in Section~\ref{sec:mpmodel} by introducing a metapopulation framework that models the spread of the disease. Building on this, Section~\ref{sec:coarse-graining} details our coarse-graining approach, a method used to simplify intricate urban structures into a synthetic hub--leaf metapopulation model. In Section~\ref{sec:def_ev}, we define epidemic vulnerability using a critical matrix that maps all possible infection pathways between hub and leaf populations. Subsequently, Section~\ref{sec:spat_temp} explores how this vulnerability is influenced by both spatial factors (the distribution of human and vector populations) and temporal factors (the patterns of mobility between locations). Finally, in Section~\ref{sec:disc} we discuss the implications of these results for extending minimal reductions to more complex metapopulation settings and for informing intervention strategies.


\section{Metapopulation model}
\label{sec:mpmodel}

\begin{figure*}
    \centering
    \includegraphics[width=12cm]{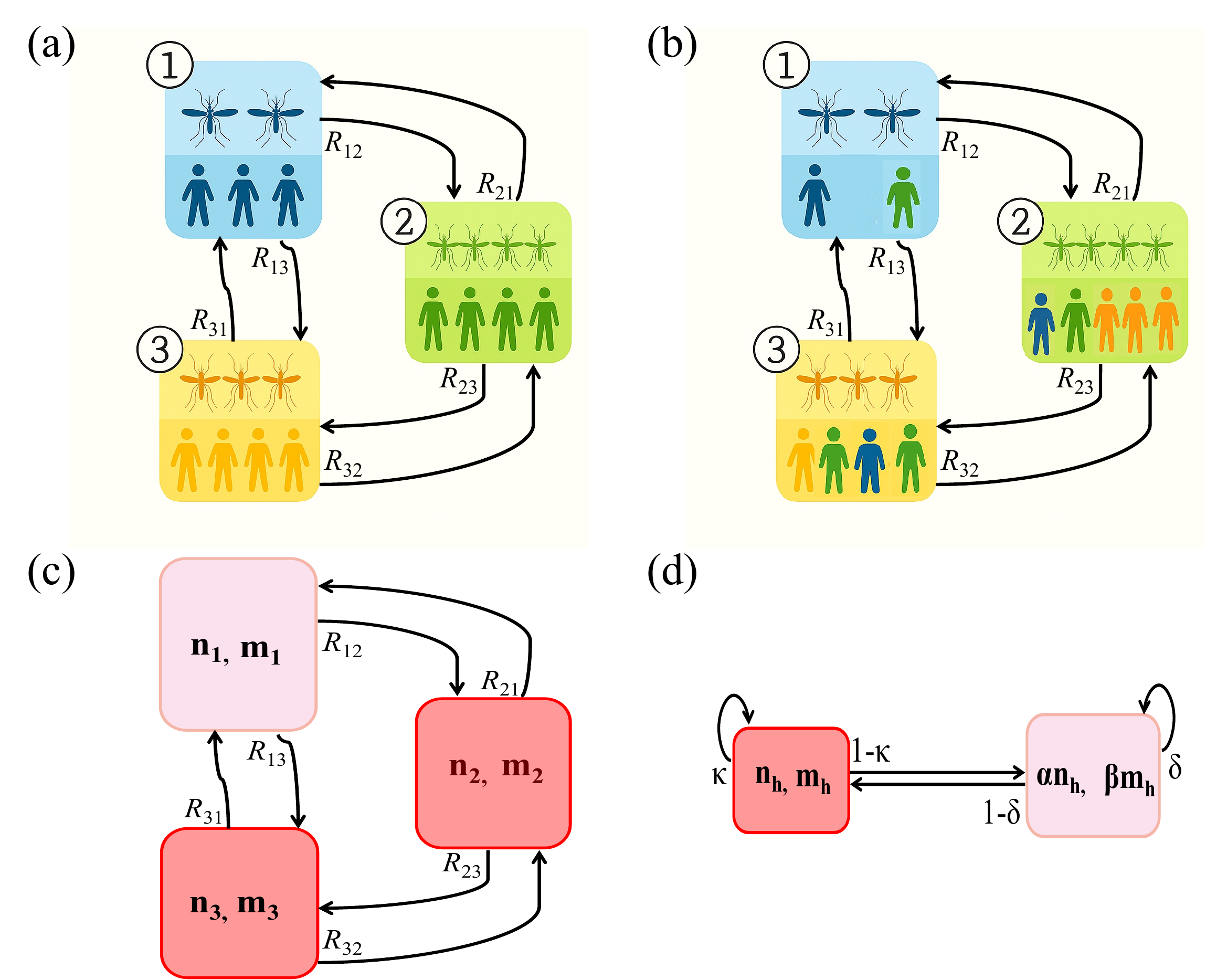}
    \caption{Model schematic, epidemic threshold, and spatial aggregation. 
Panels (a) and (b) illustrate the three-patch metapopulation framework: each node $i$ contains $n_i$ humans and $m_i$ vectors, with vectors confined to their resident node. Human mobility, described by the matrix $\mathbf{R}$, redistributes populations across patches, producing the mixing shown in (b). Ross--Macdonald dynamics then govern interactions between humans and vectors within each patch. Panels (c) and (d) show the coarse-graining procedure used to reduce complex urban networks to a hub--leaf model. High-density ``hotspot'' nodes and lower-density ``suburbs'' are first dichotomized (c), for example using the LouBar method, and then aggregated into a two-patch system (d). In this reduced model, the human and vector populations of the leaf are defined as proportions, $\alpha$ and $\beta$, respectively, of the hub's corresponding populations. Mobility parameters are also aggregated: $\kappa$ denotes the fraction of time hub residents spend within the hub (with $1-\kappa$ spent in the leaf), while $\delta$ denotes the fraction of time leaf residents spend within the leaf (with $1-\delta$ spent in the hub).}

    \label{fig:figure1}
\end{figure*}

To model the transmission of vector-borne diseases (VBDs) in a spatially structured population, we employ the Ross--Macdonald (RM) framework \cite{ross1910prevention,macdonald1957epidemiology,smith2012ross,auger2008ross,xiao2014transmission,soriano2020vector}. This approach aligns naturally with metapopulation theory, which accounts for both the spatial distribution of populations and the movement of individuals between locations. In this setting, the metapopulation is represented as a network of interconnected patches, where each patch $i$ corresponds to a geographic area. Each patch is characterized by its resident human population size ($n_i$) and vector population size ($m_i$), which can vary substantially across the network, as illustrated in Fig.~\ref{fig:figure1}(a). Humans and vectors are assigned to their home patches according to their primary residence. Human movement between patches is captured by a directed and weighted mobility matrix $\mathbf{R}$, where $R_{ij}$ denotes the fraction of individuals from patch $i$ who travel to patch $j$ in a given period. Fig.~\ref{fig:figure1}(b) illustrates how this mobility redistributes human presence across nodes. In contrast, consistent with the limited dispersal capabilities of many vector species, vector populations remain confined to their resident patches \cite{muir1998aedes}.

The probability that a human residing in location $i$ is infected at time $t$, denoted $\rho_i^H(t)$, evolves according to Markovian equations as detailed in \cite{soriano2020vector}. A central concept in this framework is the epidemic threshold: the set of critical conditions that separate a disease-free or low-level endemic state from a persistent epidemic state. To analyze this, we consider the stationary state where the infection probability remains constant over time, i.e., $\rho_i^H(t) = \rho_i^H(t+1) = \rho_i^H$. The stationary infection probability in humans, $\rho^H$, depends on the contagion rate $\lambda$. When $\lambda$ lies below a critical value---the epidemic threshold---the stationary infection probability is zero, indicating that the disease cannot sustain itself in the population. Once $\lambda$ surpasses this threshold, a bifurcation occurs and $\rho^H$ grows monotonically with $\lambda$, marking the onset of an epidemic state in which a substantial fraction of the human population remains infected.

To determine the epidemic threshold, we consider the early stage where infection prevalence is vanishingly small ($\rho_i^H \ll 1$). Under this condition, the infectious fractions of humans ($\epsilon_i^H$) and vectors ($\epsilon_i^M$) in patch $i$ follow the linearized dynamics
\begin{align}
\epsilon^H_i &= \sum_{j=1}^{N}\frac{\lambda^{MH}\beta^*}{\mu_H}
\left( pR_{ij}\frac{m_j}{n^{\mathrm{eff}}_j}
      + (1-p)\delta_{ij}\frac{m_i}{n^{\mathrm{eff}}_i} \right)
\epsilon^M_j, \label{eq:1} \\[6pt]
\epsilon^M_i &= \sum_{j=1}^{N}\frac{\lambda^{HM}\beta^*}{\mu_M}
\left( \alpha^*pR_{ij}\frac{n_j}{n^{\mathrm{eff}}_i}
      + (1-\alpha^*p)\delta_{ij}\frac{n_i}{n^{\mathrm{eff}}_i} \right)
\epsilon^H_j. \label{eq:2}
\end{align}

Here, the biological parameters are $\lambda^{MH}$ and $\lambda^{HM}$ (vector-to-human and human-to-vector transmission rates), $\beta^*$ (vector feeding rate), and $\mu_H$, $\mu_M$ (human recovery and vector mortality rates). The structural parameters are $p$ (probability of human movement), $m_j$ (vector population in patch $j$), $n_j^{\mathrm{eff}}$ (effective human population in patch $j$), and $\alpha^*$ (mobility scaling for infected humans). We emphasize that $\beta^*$ and $\alpha^*$ are distinct from the hub--leaf ratios $\alpha$ and $\beta$ introduced later.

Eqs.~(\ref{eq:1})--(\ref{eq:2}) highlight the bipartite transmission structure, with the vector-to-human and human-to-vector kernels given by
\begin{align}
M_{ij} &= pR_{ij}\frac{m_j}{n^{\mathrm{eff}}_j}
       + (1-p)\delta_{ij}\frac{m_i}{n^{\mathrm{eff}}_i}, \label{eq:3} \\[4pt]
\Tilde{M}_{ij} &= \alpha^*pR_{ji}\frac{n_j}{n^{\mathrm{eff}}_i}
               + (1-\alpha^*p)\delta_{ij}\frac{n_i}{n^{\mathrm{eff}}_i}.
\label{eq:4}
\end{align}
Note that $R_{ji}$ in $\Tilde{M}_{ij}$ accounts for the flow of humans into patch $i$ from patch $j$.

\begin{align}
\mathbf{C} = \frac{m_h}{n_h}
\begin{pmatrix}
\begin{aligned}
&\tfrac{\kappa^2}{(\kappa+\alpha(1-\delta))^2} \\
&+ \tfrac{\beta(1-\kappa)^2}{((1-\kappa)+\alpha\delta)^2}
\end{aligned}
&
\begin{aligned}
&\tfrac{\alpha\kappa(1-\delta)}{(\kappa+\alpha(1-\delta))^2} \\
&+ \tfrac{\beta\alpha(1-\kappa)\delta}{((1-\kappa)+\alpha\delta)^2}
\end{aligned}
\\[14pt]
\begin{aligned}
&\tfrac{\kappa(1-\delta)}{(\kappa+\alpha(1-\delta))^2} \\
&+ \tfrac{\beta(1-\kappa)\delta}{((1-\kappa)+\alpha\delta)^2}
\end{aligned}
&
\begin{aligned}
&\tfrac{\alpha(1-\delta)^2}{(\kappa+\alpha(1-\delta))^2} \\
&+ \tfrac{\beta\alpha\delta^2}{((1-\kappa)+\alpha\delta)^2}
\end{aligned}
\end{pmatrix}.
\label{eq:9}
\end{align}

Nontrivial solutions for $\epsilon^H$ correspond to the eigenvectors of $\mathbf{M\Tilde{M}}$, so the epidemic threshold is governed by the critical matrix $\mathbf{C = M\Tilde{M}}$. To focus on patch-level mobility, we set $p = 1$ and $\alpha^* = 1$, giving
\begin{align}
C_{ij} &= n_j \sum_{l=1}^{N} R_{il} R_{jl} \frac{m_l}{\big(n^{\mathrm{eff}}_l\big)^2},
\label{eq:5}
\end{align}
where the effective population in patch $i$ is
\begin{align}
n^{\mathrm{eff}}_i &= \sum_{j=1}^{N} n_j R_{ji}.
\label{eq:6}
\end{align}
This definition reflects the total number of individuals present in patch $i$, regardless of their origin. Eq.~(\ref{eq:5}) thus encodes all infection pathways from humans in patch $j$ to humans in patch $i$. To keep track of the notation used throughout the paper, Table~\ref{tab:glossary} summarizes the key parameters and their interpretations.

\begin{table}[ht!]
\caption{Glossary of key model parameters.}
\centering
\begin{tabular}{ll}
\hline
Symbol & Definition \\
\hline
$\alpha$ & Ratio of leaf to hub human populations \\
$\beta$ & Ratio of leaf to hub vector populations \\
$\kappa$ & Fraction of time hub residents spend in hub \\
$\delta$ & Fraction of time leaf residents spend in leaf \\
$n^{\mathrm{eff}}_i$ & Effective human population in patch $i$ \\
$\mathbf{C}$ & Critical matrix encoding infection pathways \\
$\nu$ & Epidemic vulnerability \\
\hline
\end{tabular}
\label{tab:glossary}
\end{table}

\section{Coarse-grained synthetic hub--leaf metapopulation model}
\label{sec:coarse-graining}

The coarse-graining approach used to simplify complex urban environments for VBD transmission analysis is shown in Fig.~\ref{fig:figure1}(c,d). Panel (c) depicts the initial dichotomization of nodes in a real urban network into high-density ``hotspots'' and lower-density ``suburbs,'' often identified using the LouBar method \cite{li2018effect,bassolas2019hierarchical}.  

Panel (d) then shows the aggregation of these two classes into a synthetic hub--leaf metapopulation model. The hub represents the combined hotspots, while the leaf represents the aggregated suburbs. In this reduced system, the leaf's human population is a proportion $\alpha$ of the hub's, and its vector population is a proportion $\beta$ of the hub's vector population. Mobility is likewise parameterized: $\kappa$ denotes the fraction of time hub residents spend in the hub (with $1-\kappa$ in the leaf), while $\delta$ denotes the fraction of time leaf residents spend in the leaf (with $1-\delta$ in the hub).  

This reduction systematically condenses the complexity of real urban networks into a tractable two-patch system. By focusing on the essential interplay between dense hubs and sparse leaves, we can isolate the universal factors that govern the vulnerability. The hub--leaf framework therefore provides both a conceptual and computational platform for exploring parameter space and identifying the drivers that amplify or suppress VBD outbreaks in interconnected populations.

For the simplified two-patch case (hub $i$ and leaf $j$), the effective populations are
\begin{align}
n^{\mathrm{eff}}_i &= n_iR_{ii}+n_jR_{ji}, \label{eq:7a}\\
n^{\mathrm{eff}}_j &= n_iR_{ij}+n_jR_{jj}. \label{eq:7b}
\end{align}
Substituting these into Eq.~(\ref{eq:5}), the critical matrix for the two-patch metapopulation can be written compactly as
\begin{align}
\mathbf{C} = \frac{m_i}{n_i}\Bigg[ &
 \frac{1}{\big(R_{ii}+\tfrac{n_j}{n_i}R_{ji}\big)^2}
 \begin{pmatrix}
R_{ii}^2 & \tfrac{n_j}{n_i}R_{ii}R_{ji}\\[6pt]
R_{ii}R_{ji} & \tfrac{n_j}{n_i}R_{ji}^2
\end{pmatrix} \notag \\[8pt]
&+ \frac{\tfrac{m_j}{m_i}}{\big(R_{ij}+\tfrac{n_j}{n_i}R_{jj}\big)^2}
\begin{pmatrix}
R_{ij}^2 & \tfrac{n_j}{n_i}R_{ij}R_{jj}\\[6pt]
R_{ij}R_{jj} & \tfrac{n_j}{n_i}R_{jj}^2
\end{pmatrix}\Bigg].
\label{eq:8}
\end{align}
For our one-hub--leaf model [Fig.~\ref{fig:figure1}(d)], we define the parameters as follows: the hub, designated as patch $i$, has a human population $n_h$ and a vector population $m_h$. For its residents, the fraction of time spent within the hub is $R_{ii}=\kappa$, while the fraction of time spent in the leaf is $R_{ij}=1-\kappa$. Similarly, for the leaf, designated as patch $j$, the fraction of time its residents spend in the leaf is $R_{jj}=\delta$, and the fraction of time spent in the hub is $R_{ji}=1-\delta$. We further define $\alpha$ as the ratio of the leaf to hub human populations ($n_l=\alpha n_h$) and $\beta$ as the ratio of the leaf to hub vector populations ($m_l=\beta m_h$). These scaling factors $\alpha$ and $\beta$ are distinct from the behavioral and biological parameters $\alpha^*$ and $\beta^*$ introduced earlier.

In this representation, $\mathbf{C}$ encodes all infection pathways between hub and leaf. The spectral radius of $\mathbf{C}$ (its largest eigenvalue) determines the epidemic threshold: values below one correspond to self-limiting outbreaks, whereas values exceeding one signal conditions favorable to sustained transmission. In the hub--leaf model, this spectral radius integrates the effects of four key parameters: the relative human population size in the leaf ($\alpha$), the relative vector population size in the leaf ($\beta$), fraction of time hub residents stay in hub ($\kappa$), and fraction of time leaf residents stay in leaf ($\delta$). Thus, $\rho(\mathbf{C})$ serves as a compact measure of the vulnerability in this minimal two-patch system.

\section{Defining epidemic vulnerability}
\label{sec:def_ev}

\begin{figure*}
    \centering
    \includegraphics[width=14cm]{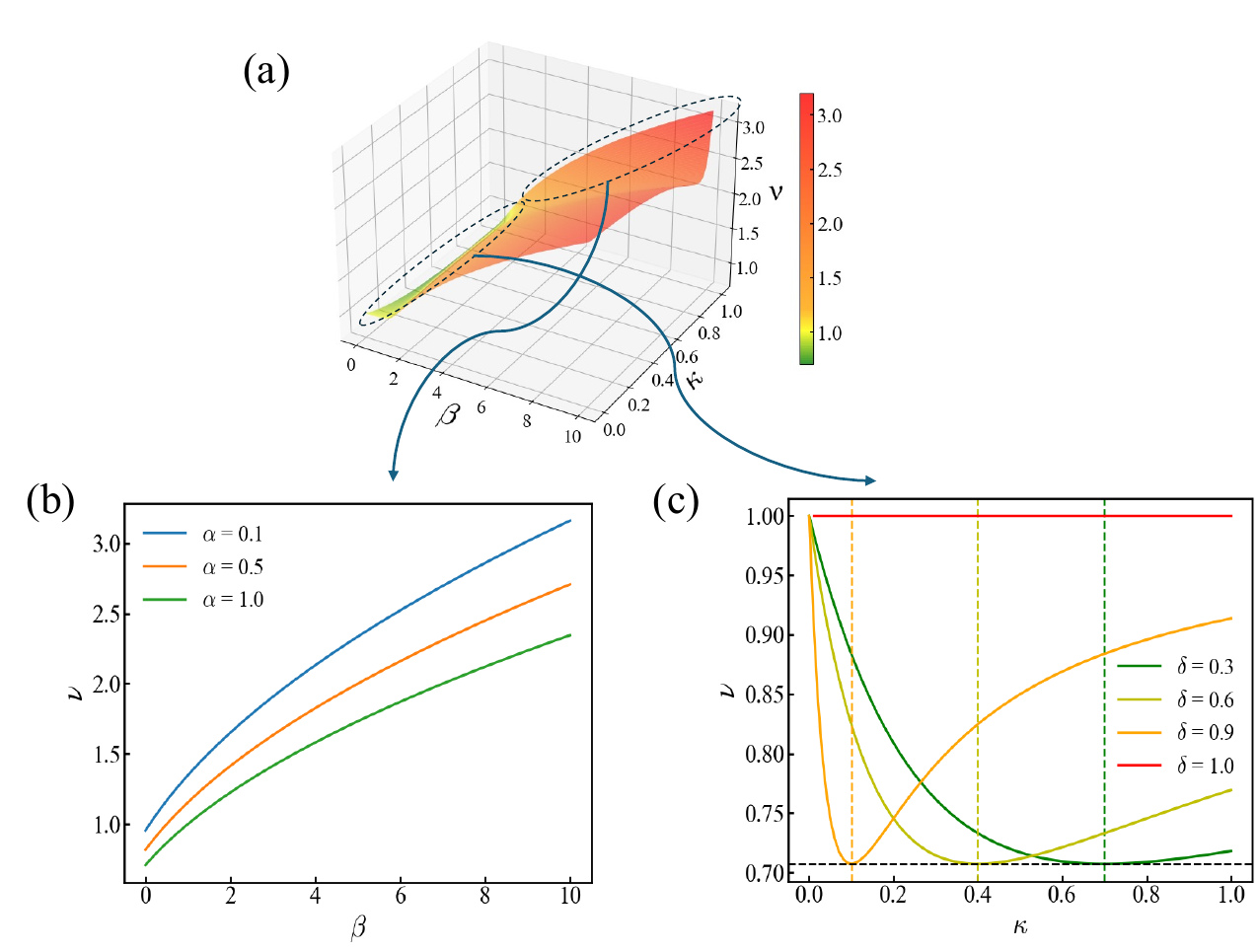}
    \caption{Spatial and temporal patterns of epidemic vulnerability. 
    (a) Vulnerability surface as a function of vector ratio $\beta$ and hub mobility $\kappa$ ($\alpha=1$, $\delta=0.1$).
    (b) Spatial slice: $\nu$ vs.~$\beta$ for different $\alpha$ under balanced mobility ($\kappa=1-\delta$).
    (c) Temporal slice: $\nu$ vs.~$\kappa$ for different $\delta$, with negligible leaf vectors ($\beta \ll 1$) and $\alpha=1$.}
    \label{fig:fig2}
\end{figure*}

Building upon the idea that the maximum eigenvalue (spectral radius) of the critical matrix $\mathbf{C}$ quantifies the epidemic vulnerability of our one-hub--leaf system to VBDs, we denote this quantity by $\nu$. The overall scale factor $\tfrac{m_h}{n_h}$ in $\mathbf{C}$ does not affect the eigenvalues, which are obtained from the characteristic polynomial. Therefore, this factor can be disregarded when calculating $\nu$. The relevant eigenvalue, determined by the human and vector populations and their mobility between hub and leaf, is given by:


\begin{align}
\nu = &\Bigg( \tfrac{1}{2}\Big[ 
   \Tilde{\Theta}_h(\kappa,\delta) + \beta \Tilde{\Theta}_l(\kappa,\delta) \notag \\
   & + \sqrt{\Big(\Tilde{\Theta}_h(\kappa,\delta) + \beta \Tilde{\Theta}_l(\kappa,\delta)\Big)^2 
   - \alpha \beta \,\Big(\Tilde{\Theta}_{hl}(\kappa,\delta)\Big)^2} \,\Big] \Bigg)^{\frac{1}{2}}.
\label{eq:7}
\end{align}
Eq.~(\ref{eq:7}) links system vulnerability to the key spatiotemporal parameters of the model: $\kappa$ (fraction of time hub residents stay in hub), $\delta$ (fraction of time leaf residents stay in leaf), $\alpha$ (relative human population in the leaf), and $\beta$ (relative vector population in the leaf). The vulnerability $\nu$ is composed of three components, each depending on $\kappa$ and $\delta$:
\begin{subequations}\label{eq:theta_defs}
\begin{align}
\Tilde{\Theta}_h(\kappa,\delta) &=
  \tfrac{\kappa^2+\alpha(1-\delta)^2}{\big(\kappa+\alpha(1-\delta)\big)^2},\\
\Tilde{\Theta}_l(\kappa,\delta) &=
  \tfrac{(1-\kappa)^2+\alpha\delta^2}{\big((1-\kappa)+\alpha\delta\big)^2},\\
\Tilde{\Theta}_{hl}(\kappa,\delta) &=
  \tfrac{2(1-\kappa-\delta)}{\big(\kappa+\alpha(1-\delta)\big)\big((1-\kappa)+\alpha\delta\big)},
\end{align}
\end{subequations}
where the subscripts $h,l,hl$ refer to hub, leaf and cross-vulnerability respectively.

\begin{figure*}[t!]
    \centering
    \includegraphics[width=16cm]{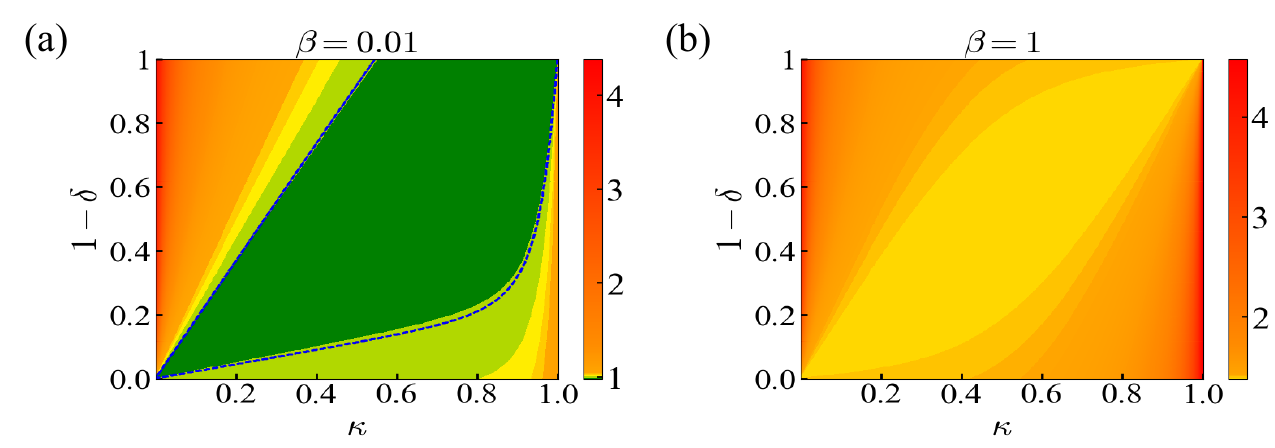}
    \caption{Contour plots of epidemic vulnerability $\nu$ determined by mobility parameters, for different $\beta$ values at fixed $\alpha=0.05$. Larger $\beta$ increases overall vulnerability. The blue dotted contour line indicates the threshold where vulnerability equals 1. Color scales are similar across panels but not identical.}

    \label{fig:3}
\end{figure*}

A defining feature of VBD transmission is the localized nature of vector--host interactions. Vectors remain confined to their patch and interact only with humans present there. Unlike directly transmitted diseases, where mobility drives spread between locations, vulnerability within a patch persists as long as vectors are present. Mathematically, this leads to a contact term proportional to $\tfrac{1}{(n^{\mathrm{eff}})^2}$ in the critical matrix that underlies Eq.~(\ref{eq:7}).

For the hub vulnerability $\Tilde{\Theta}_h(\kappa,\delta)$, the minimum value $\tfrac{1}{1+\alpha}$ occurs when $\kappa=1-\delta$, corresponding to balanced flows between hub and leaf. The maximum approaches $\tfrac{1}{\alpha}$ (for $\alpha \leq 1$) when hub mobility is low ($\kappa \approx 0$) and leaf mobility allows inflow ($\delta < 1$), reducing the effective hub population.

The leaf vulnerability $\Tilde{\Theta}_l(\kappa,\delta)$ also reaches a minimum of $\tfrac{1}{1+\alpha}$ under balanced flows ($\kappa=1-\delta$). Its maximum approaches $\tfrac{1}{\alpha}$ (for $\alpha \leq 1$) when hub mobility is high ($\kappa \approx 1$) and $\delta>0$, maintaining a resident population in the leaf. The contribution of the leaf is scaled by $\beta$, highlighting the role of vector abundance in shaping overall vulnerability.

The cross-vulnerability term $\Tilde{\Theta}_{hl}(\kappa,\delta)$ measures interdependence between hub and leaf due to mobility. It vanishes when $\delta=1-\kappa$, again indicating balanced flows. Its maximum, $\tfrac{2}{\alpha}$ (for $\alpha \leq 1$), occurs when both mobilities are low ($\kappa \approx 0$, $\delta \approx 0$), corresponding to high interpatch mixing. However, since this term is scaled by $\alpha\beta$, its influence is typically smaller than the direct hub and leaf terms.
By analyzing Eq.~(\ref{eq:7}) and its dependence on $\alpha$, $\beta$, $\kappa$, and $\delta$, we can identify the parameters that most strongly amplify or mitigate the risk of VBD outbreaks in the hub--leaf system.

\section{Spatiotemporal determinants}
\label{sec:spat_temp}

Epidemic vulnerability in VBD depends jointly on spatial factors---the relative distribution of human and vector populations---and temporal factors governing mobility between locations. Spatial parameters, represented by the human ($\alpha$) and mosquito ($\beta$) population ratios between hub and leaf, determine transmission potential by setting local host--vector densities. Temporal parameters, captured by hub and leaf mobility values ($\kappa$ and $\delta$), regulate the allocation of time across patches, shaping opportunities for exposure and transmission.  

Fig.~\ref{fig:fig2}(a) illustrates these interactions by plotting epidemic vulnerability $\nu$ as a function of $\beta$ and $\kappa$, for $\alpha=1$ and $\delta=0.1$. The resulting surface highlights the difficulty of isolating individual parameter effects in high-dimensional space, where spatial and temporal factors combine nonlinearly to determine overall system vulnerability.

We study two slices of the parameter space in Fig.~\ref{fig:fig2}(a). First, under balanced mobility ($\kappa=1-\delta$ with $\kappa=0.9$ as $\delta=0.1$), we isolate spatial effects. Second, under minimal vector presence ($\beta \ll 1$) and a balanced human population ($\alpha=1$), we isolate temporal effects. Together these scenarios reveal the separate and combined roles of spatial and temporal factors in shaping vulnerability.

\subsection{Spatial drivers}

We first analyze the influence of spatial parameters under balanced mobility, $\kappa=1-\delta$. As established earlier, this condition minimizes patch vulnerabilities and removes the cross-vulnerability term, decoupling hub and leaf contributions. Under this simplification, Eq.~(\ref{eq:7}) reduces to
\begin{equation}
    \nu = \sqrt{\Tilde{\Theta}_h + \beta\Tilde{\Theta}_l} 
        = \sqrt{\frac{1}{1+\alpha} + \beta\frac{1}{1+\alpha}} 
        = \sqrt{\frac{1+\beta}{1+\alpha}}.
    \label{eq:8}
\end{equation}
Here vulnerability depends only on $\alpha$ and $\beta$, the relative human and vector populations.  

Fig.~\ref{fig:fig2}(b) illustrates $\nu$ as a function of $\beta$ for fixed $\alpha$. As $\beta$ increases, $\nu$ rises monotonically: a larger fraction of vectors in the leaf amplifies system-wide risk. The human ratio $\alpha$ modulates this relationship. Smaller $\alpha$ (fewer humans in the leaf relative to the hub) yields higher vulnerability, while larger $\alpha$ reduces vulnerability by diluting per-vector contact rates.  

The effect of $\beta$ is nonlinear: $\nu$ increases sharply for small $\beta$, then levels off. This means that even modest initial shifts of vectors toward the leaf substantially elevate risk, while further increases have diminishing effect. In sum, spatial heterogeneity in host and vector distributions strongly shapes vulnerability: more vectors in the leaf consistently raise risk, but more humans there mitigate it.

\begin{figure*}[ht!]
    \centering
    \includegraphics[width=16cm]{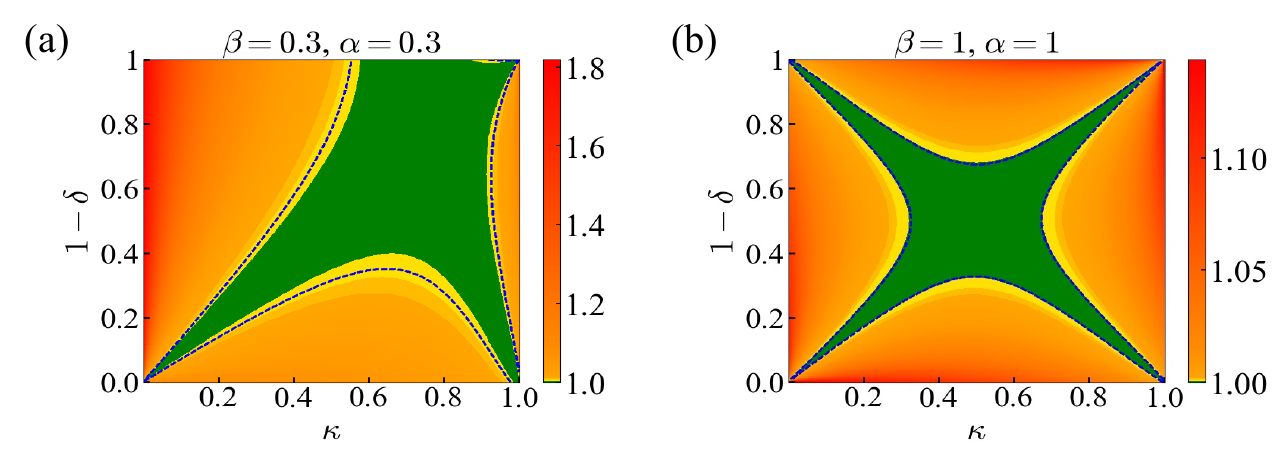}
    \caption{Contour plots of epidemic vulnerability $\nu$ for $\beta/\alpha \approx 1$, showing how balanced vector-to-human ratios in hub and leaf shape mobility-driven dynamics. The blue dotted contour line indicates the threshold where vulnerability equals 1. Color scales are similar across panels but not identical.}

    \label{fig:4}
\end{figure*}

\subsection{Temporal drivers}

Next we examine temporal dynamics by focusing on mobility parameters $\kappa$ and $\delta$ while suppressing spatial effects. To do so, we set $\beta \ll 1$ (negligible leaf vectors) and $\alpha=1$ (balanced human populations). Under this assumption, the vulnerability simplifies to
\begin{equation}
    \nu \approx \sqrt{\frac{\kappa^2+\alpha(1-\delta)^2}{\big(\kappa+\alpha(1-\delta)\big)^2}}.
    \label{eq:9}
\end{equation}
Fig.~\ref{fig:fig2}(c) shows $\nu$ as a function of $\kappa$ for different $\delta$, revealing three temporal phases:
\begin{enumerate}
    \item Low hub mobility ($\kappa \approx 0$): Vulnerability is $\nu \approx 1$, dominated by incoming leaf residents. Decreasing $\delta$ increases their time in the hub but also dilutes exposure, keeping vulnerability roughly constant.
    \item Intermediate $\kappa$: As hub residents spend more time locally, external exposure decreases, producing an inverse relation between $\kappa$ and $\nu$. This effect is strongest at small $\delta$ (significant outflow from the leaf). When $1-\delta \gtrsim \kappa$ (i.e., $\kappa+\delta \lesssim 1$), leaf inflow dominates hub numbers, reinforcing this decline.
    \item High leaf mobility ($\delta=1$): With no outflow from the leaf, hub and leaf decouple. Vulnerability stabilizes at $\nu=1$, independent of $\kappa$.
\end{enumerate}
Thus, temporal vulnerability evolves through phases shaped by hub mobility, leaf mobility, and their interplay. Minimal leaf vectors confine risk to the hub, but mobility flows strongly modulate the magnitude of this vulnerability.

\subsection{Coupled spatiotemporal drivers}

The preceding analyses showed that spatial parameters ($\alpha$, $\beta$) and temporal mobility parameters ($\kappa$, $\delta$) exert distinct influences on epidemic vulnerability. Yet in realistic settings these drivers act jointly. The relative abundance of humans ($\alpha$) and vectors ($\beta$) in the leaf sets its baseline susceptibility, while mobility patterns ($\kappa$, $\delta$) govern mixing with the hub. Their interplay redistributes the contributions of hub and leaf to overall vulnerability, producing qualitatively different regimes. To explore these systematically, we analyze three representative cases: (i) $\alpha\beta \ll 1$ (at least one leaf population is small), (ii) $\beta/\alpha \approx 1$ (vectors and humans scale proportionally across hub and leaf), and (iii) $\beta/\alpha \gg 1$ (the leaf vector population is disproportionately large).

\begin{figure*}
    \centering
    \includegraphics[width=16cm]{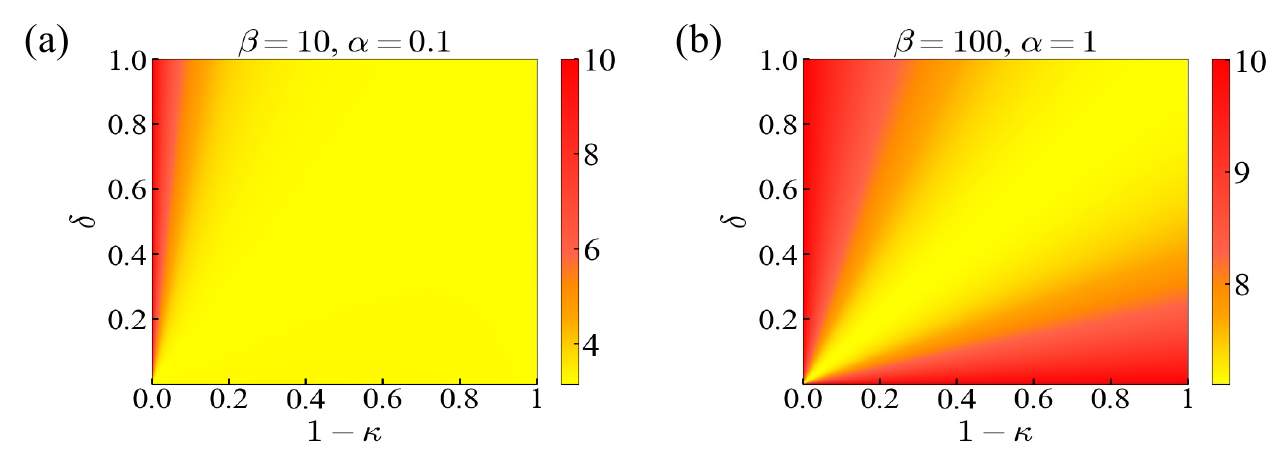}
   \caption{Contour plots of epidemic vulnerability $\nu$ for $\beta/\alpha \gg 1$. Axes switched ($x$: $1-\kappa$, $y$: $\delta$) to highlight leaf-centric dynamics. Color scales are similar across panels but not identical.}
    \label{fig:5}
\end{figure*}

\subsubsection{$\alpha\beta \ll 1$}

Building on the temporal analysis where $\beta \ll 1$ led to hub-centric vulnerability, we now examine the regime where $\beta$ is significant but $\alpha \ll 1$, under the constraint $\alpha\beta \ll 1$. In this context, since the leaf's population is small relative to the hub's, the vulnerability is predominantly influenced by the hub's mobility ($\kappa$) and the residence time of its population in each patch. The resulting vulnerability is

\begin{equation}
    \nu = \sqrt{\frac{\kappa^2+\alpha(1-\delta)^2}{\big(\kappa+\alpha(1-\delta)\big)^2} + \frac{\beta(1-\kappa)^2}{\big((1-\kappa)+\alpha\delta\big)^2}}.
    \label{eq:15}
\end{equation}

As in the temporal analysis, three regimes emerge: (1) low $\kappa$, (2) $(1-\delta)\gtrsim\kappa$, and (3) $\kappa \gg 1-\delta$. A key difference arises in regime (3): when $\beta \ll 1$, vulnerability was capped at 1, but as $\beta$ increases, the leaf contribution elevates baseline vulnerability and broadens this regime. At $\beta=1$, the phase diagram becomes nearly symmetrical, with low- and high-$\kappa$ regimes separated by a transition near $\kappa \approx 1-\delta$.

\subsubsection{$\beta/\alpha \approx 1$}
When $\beta/\alpha \approx 1$, vector-to-human ratios are similar in hub and leaf ($m_l/n_l \approx m_h/n_h$). This produces near-symmetric contagion dynamics. With small $\alpha$ and $\beta$ (e.g., 0.3 each), vulnerability is minimized at high $\kappa$ and $1-\delta$, since additional time in the hub dilutes exposure. As $\alpha$ and $\beta$ increase toward unity, both populations become equally vulnerable, requiring more balanced mobility to minimize risk. Across parameter space, vulnerability consistently reaches a minimum around $\kappa = 1-\delta$, reflecting a compensatory mechanism where outflow from one patch balances the lack of outflow from the other, enhancing mixing and reducing risk.

\subsubsection{$\beta/\alpha \gg 1$}

When the leaf hosts a disproportionately large vector population ($\beta/\alpha \gg 1$), vulnerability becomes leaf-dominated. Here the dynamics mirror the hub-centric analysis but inverted. At low hub-to-leaf influx ($1-\kappa$), the small resident leaf population is heavily exposed to dense vectors. As $1-\kappa$ increases, incoming hub residents dilute this exposure, reducing vulnerability. Internal leaf mobility $\delta$ modulates the effect: low $\delta$ disperses contact, mitigating the risk. Conversely, high $\delta$ prolongs exposure within the leaf, which amplifies the risk. The net result is a vulnerability landscape dominated by leaf dynamics, with overall risk elevated relative to the hub-centric case.

Taken together, these regimes reveal how scaling the human and vector populations between hub and leaf shifts vulnerability from hub-dominated to symmetric to leaf-dominated dynamics. The mobility parameters $\kappa$ and $\delta$ govern the transitions, with $\kappa = 1-\delta$ consistently marking conditions of minimized vulnerability through compensatory mixing. This analysis highlights how joint scaling of populations and mobility generates the spatiotemporal structure of epidemic risk in VBD systems.

\section {Conclusion}
\label{sec:disc}

This study has systematically identified the spatiotemporal determinants of vector-borne disease (VBD) outbreaks in interconnected metapopulation through a coarse-grained hub--leaf model. By deriving and analyzing the epidemic vulnerability equation, we showed how the relative sizes of human ($\alpha$) and vector ($\beta$) populations, together with temporal mobility parameters ($\kappa$ and $\delta$), govern the system's susceptibility to outbreaks. Specific parameter regimes, such as $\alpha\beta \ll 1$ and $\beta/\alpha \approx 1$, highlighted the shifting balance of vulnerability between hub and leaf populations, while the condition $\kappa = 1-\delta$ consistently minimized risk by compensating flows across patches. In contrast, the $\beta/\alpha \gg 1$ regime demonstrated how a disproportionately large vector population can make the leaf the dominant center of vulnerability.

Beyond these analytical results, our framework advances a more general perspective on epidemic risk in spatially structured populations. The hub--leaf model isolates universal principles that extend beyond its immediate setting. For example, maximum entropy approaches have been used to infer spatial sources of disease outbreaks \cite{Ansari_2022,hampton2011adjusting,angulo2013spatiotemporal}, illustrating how statistical inference can complement mechanistic models in identifying where vulnerability concentrates within heterogeneous landscapes. Incorporating such methods into the present framework opens a path toward predictive tools that integrate mechanistic dynamics with data-driven inference.

Looking ahead, several directions emerge. First, extending the model to include finer-grained demographic or spatial heterogeneity, seasonal forcing, and dynamic behavioral responses would make it applicable to real-world systems with greater fidelity \cite{rodriguez2001models,favier2005influence,altizer2006seasonality,chen2018feedback}. Second, embedding the framework within multi-patch or network models could reveal how vulnerability scales across hierarchies of urban or regional connectivity \cite{hazarie2021interplay, soriano2022modeling,pardo2023epidemic}. Third, incorporating socioeconomic and environmental covariates would allow exploration of how structural inequalities and climate variability interact with host--vector dynamics to amplify or mitigate risk \cite{shocket2020environmental,athni2021influence}. Finally, linking these theoretical advances to empirical surveillance data offers the prospect of developing real-time indicators of vulnerability, enabling targeted and adaptive public health interventions \cite{zeng2021artificial,jia2023innovations,nuha2025beyond}.

In sum, this work provides a conceptual and mathematical foundation for understanding how spatial distributions and temporal mobilities jointly shape epidemic vulnerability. By combining coarse-grained analytical insight with complementary inference methods and extending into richer multi-scale models, one can move toward a general framework for identifying and mitigating epidemic risk in complex, interconnected populations.

\section*{Acknowledgements}
GG acknowledges partial support through Grant No. 62417 from the John Templeton Foundation. The opinions expressed in this publication are those of the author(s) and do not necessarily reflect the views of the Foundation. GG and BP also acknowledge support from the University of Rochester. We thank David Soriano Pan\~os for useful discussions.
\nocite{*}

\bibliography{apssamp-1}

\end{document}